# A Functional Model For Information Exploration Systems

THIAGO NUNES, Instituto Federal Fluminense

DANIEL SCHWABE, Pontifical Catholic University of Rio de Janeiro

## Abstract

Information exploration tasks are inherently complex, ill-structured, and involve sequences of actions usually spread over many sessions. When exploring a dataset, users tend to experiment higher degrees of uncertainty, mostly raised by knowledge gaps concerning the information sources, the task, and the efficiency of the chosen exploration actions, strategies, and tools in supporting the task solution process. Provided these concerns, exploration tools should be designed with the goal of leveraging the mapping between user's cognitive actions and solution strategies onto the current systems' operations. However, state-of-the-art systems fail in providing an expressive set of operations that covers a wide range of exploration problems. There is not a common understanding of neither which operators are required nor in which ways they can be used by explorers. In order to mitigate these shortcomings, this work presents a formal framework of exploration operations expressive enough to describe at least the majority of state-of-the-art exploration interfaces and tasks. We also show how the framework leveraged a new evaluation approach, where, we draw precise comparisons between tools concerning the range of exploration tasks they support.

## 1 INTRODUCTION

The last decade has experienced an exponential growth of the amount of publicly available data, mostly leveraged by the spreading of the WWW, the development of semantic technologies, and the Linked Data community[1]. Since the mass publication of semi-structured information has taken place, the search for information has been one of the most relevant themes in information systems research. Unfortunately the development of efficient applications to empower the searcher in finding and processing relevant pieces of information has not evolved with the same speed, leaving large parts of the published data effectively out of reach.

It has been a consensus among the research community that traditional Information Retrieval (IR) tools are not sufficient in supporting data exploration behaviors [8,27,28,39], which take place when the user lacks the knowledge and ability to precisely describe the characteristics of the desired items in terms of keywords. The interaction in IR tools, such as Google, and Yahoo!, is based on isolated sequences of query-response actions, where the user formulates a query and the system returns a set of documents that match the query. Despite the simplicity of use of the query-response model, it has been strongly criticized with regards to supporting more complex information search tasks, such as Exploratory Searches [28]. As an example, a searcher can solve the task "discover who invented the light bulb" in a single query against search engines. On the other hand, the task "write a paper about recently discovered treatments for diabetes" would require exploratory behavior and, consequently, more advanced support for browsing, filtering, aggregating, and comparing information items. In fact, traditional IR designers have already recognized the need to support more complex information searches and question-answering tasks in their tools. For example, Google and Microsoft search engines already provide result set browsing and filtering operations based on semi-structured data, leveraged by their knowledge graphs [36]. The support for exploration tasks, though, is still in its infancy.

---

[0]

[1] http://linkeddata.org/

Even though there has been much work on the development of computational systems supporting exploration tasks, visualization systems, faceted search tools, and set-oriented interfaces, the lack of a formal understanding of the exploration process and its operations and the absence of a proper separation of concerns approach in the design phase is the cause of many expressivity issues and serious limitations. The exploration environments usually do not present precise descriptions of neither what operations they support nor what combinations of operators are available through their interfaces. Thus, it is not possible to assess how good is the support of an environment for a given exploration task from the functional perspective. As a consequence, no matter how advanced the interface design is, a missing operator may cause serious limitations to the task resolution process and the range of distinct solution strategies the explorer can pursue.

This work presents a novel formalization of exploration processes and operations that is able to describe the majority of published exploration environments and tasks types reported in the state-of-the-art literature. The evaluation of the proposed framework is guided by case studies and descriptions of state-of-the-art tools. Moreover, we present a new approach to evaluate the expressivity of exploration tools, where we can more precisely assess the extent of their support for exploration tasks. We describe the possible uses and compositions of exploration operators in each tool by analyzing their tactical and strategic support, abstracting interface and interaction issues. The results show the relevance of the presented framework for describing exploration tasks and solution processes, as well as, assessing and comparing tools from the expressivity perspective.

The organization of this work is as follows. Section 2 presents a conceptual view of the architecture of an exploration tool and the formal description of the proposed framework of exploration operations. Section 3 presents a case study evaluation of the framework. In section 4, we discuss a new evaluation approach leveraged by the framework of operations and the separation of functional and interface concerns. Section 5 presents implementation details of the proposed framework in the Xplain exploration environment. Section 6 presents the related works. In section 7, we present the conclusions and future works.

## 2   EXPLORATION FRAMEWORK

We propose the organization of the design space of exploration tools in a three-layer architecture: Data Model, Functional, and Interaction/Interface, shown in Fig. 1. The Data Model layer is responsible for providing access to data repositories and mapping their data models to an abstract data model, which will be manipulated by the user throughout the exploration task. The Functional layer presents a set of exploration operators whose repeated compositions capture the solution strategies adopted by the users. The Interaction/Interface layer provides proper access to both the operators and the data items being manipulated.

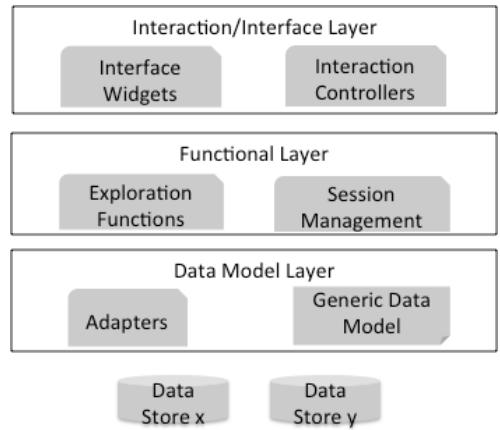

**Fig. 1. The layers of an exploration environment architecture.**

In Fig. 1, the Data Model layer addresses the design of data representation and access, which comprises a generic data model, storage and retrieval techniques. The second layer addresses the exploration functions, with their parameters and results, and the management of task progress, represented by the Session Management module. The Interaction/Interface layer addresses interaction and interface issues that support both the execution of the exploration functions and task management actions along the exploration process. The following sections discuss the responsibilities of each layer.

## 2.1    Framework Construction

The need for a common agreement on the set of operations with satisfactory expressivity for describing exploration tools has already been asserted in the literature [10,20,42], however, it remains an open question. We share the vision that achieving a complete framework, if at all possible, is a very hard task [41]. Nonetheless, the claim of this work is to describe a representative set of state-of-the-art operations, rather than all possible actions.

In order to define the framework, we first carried out a literature survey for papers presenting Exploratory Search tools or Information Exploration environments and defined the main categories of the tools, namely Faceted Search Tools, Set-Oriented Browsers, Tabular Processors, and Relation Finders. For each tool, we analyzed their interfaces (when available), bibliography, and tutorials to synthesize a collated list of features. Next, we analyzed the list to devise a set of operations and parameters, looking at the following tools: Tabulator, Mspace [34], gfacet [24], /facet [25], tfacet [13], Sewelis [20], SemFacets [7], Parallel Faceted Browser [14], Visor [32], Parallax, Rhizomer [21], MusicPinta [19], Relation Browser [43], BrowseRDF [30], Liquid Query [11], SearchComputing [10], Tableau [22], Explorator [6] and its follow up RExplorator [18], Fusion [5], and RelFinder [23].

We present below a general description of the operations that describe at least the exploration actions of these state-of-the-art exploration tools:

- *Refine(Items*, *Filter)*: filters the current set of items to a set of items that matches the restrictions imposed by the explorer through the *Filter* parameter. The *Refine* operation along with *Pivot* and set operations describe the majority of faceted search tools;

- *Group(Items*, *Relation)*: groups a set of items based on a relation, which can be defined either by the data model or by the user as a computed relation. It is important to observe that this is an abstract definition, which can be specialized, for example, by a clustering operation, where the relation is a distance function;

- *Correlate(SourceItems, TargetItems)*: finds a set of relations that connects the two sets of items. This operation describes a category of exploration tools that discovers connections between items, such as, Fusion, RelFinder, and Visor;

- *Rank(Items, Fscore)*: ranks a set of items given a score function;

- *Map(Items, MappingFunction):* maps a set of items onto another set of items using a mapping function, where, the function is provided by the environment, such as, counts, format and scale converters. This function is usually found in tabular processors for creating new columns (relations), thus extrapolating the relations described by the data schema;
- *Unite(Set1, Set2):* computes the union between two sets of items;
- *Intersect(Set1, Set2):* intersects two sets of items;
- *Diff(Set1, Set2):* computes the difference between two sets of items.

Next, we define the basic notation and the formal descriptions of the generic data model and exploration operations.

## 2.2 Basic Notation and Operators

This section presents a formalization of the exploration operations proposed. We first establish some basic notation and definitions, next we present the generic data model, and finally, the formal descriptions of the exploration process and the operations.

We define our model using set theory and logical operators, as follows. Sets can be denoted by both the enumeration of its elements, e.g. S = $\{e_1, e_2, ..., e_n\}$, and by specifying a property *P* of its members: S = { e | P(e) }. Each property is described in terms of the following logical operators:

- $\rightarrow$ stands for "implies"
- $\leftrightarrow$ stands for "if and only if"
- $\land$ stands for "and"
- $\lor$ stands for "or"
- $\neg$ stands for "not"
- $\exists$ stands for "there exists"
- $\forall$ stands for "for all"
- $\equiv$ stands for "equivalent of"

We denote $|S|$ as the number of members of a set S. In the case of partially ordered sets (POS), we refer to specific elements using indexes. Let *S* be a partially ordered set, $S_{[n]}$ refer to the nth element of the set, where *n* is an element of some indexing set over *S*.

We describe relations as ordered pairs of elements, which we denote using "<" ">", e.g. <a1,a2>, <i1,i2>. We also denote relations as subsets of the Cartesian product between sets, e.g., let A = {i1, i2} and B = {a1, a2}, the Cartesian product $A \times B = \{<i1, a1>, <i1, a2>, <i2, a1>, <i2, a2>\}$. A binary relation $\Phi \subseteq A \times B$ is a subset of their Cartesian product, whose elements $< i, j > \in \Phi$ can be denoted as $i\Phi j$. n-Cartesian products of a single set $A \times_1 ... \times_n A$ we denote as $A^n$.

We define the formalizations of each operation in terms of basic set operations, defined as follows:

*Definition 1.* The union between any two sets A and B is: $A \cup B = \{x | x \in A \lor x \in B\}$.

*Definition 2.* The intersection between any two sets A and B is: $A \cap B = \{x | x \in A \land x \in B\}$.

*Definition 3.* The difference between any two sets A and B is: $A - B = \{x | x \in A \land \neg x \in B\}$.

## 2.3 The Data Model

There are many data models described in the literature, such as the relational model [29], RDF[2], and a variety of NoSQL models [15]. Although the validity of these models has been extensively proven for their respective domains of problems, we see the exploration process as not attached to the specificities of a single model; it can be described in terms of a generic data model that can be further mapped to each one of these. Therefore, we created a simplified version of the Entity-Relationship model [16], which suffices for our purposes.

---

[2] https://www.w3.org/TR/WD-rdf-syntax-971002/

Independently of how the data is represented, the user always manipulates items and relationships among them. Items can be organized into groups, such as "papers by author" or "papers by venue". Groups can also be formed along more than one dimension, such as "papers by author by publication year". Therefore, the design solution adopted was to model items and relations as nested relations. As an example, "papers by author by publication year" becomes a three-level nesting, having the papers grouped by year inside a group for each author. Nesting relations can be represented as trees, illustrated in Fig. 2. We call "exploration set" any nested relation generated by the execution of an exploration action.

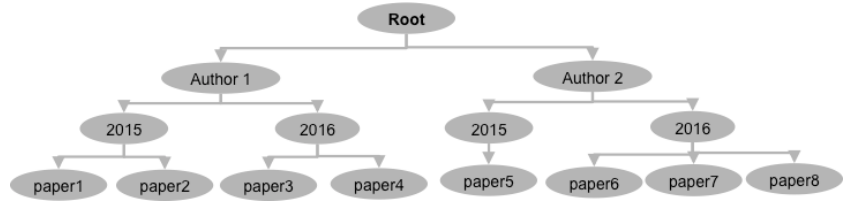

**Fig. 2. Nesting of groups of papers by author by publication year.**

The binary relation $\preccurlyeq_T$ defines the child-parent relationships of the tree and is also well founded, i.e., there is a least item $i \in t$ that has no parent, which is the root of the tree. More formally, the properties of the relation trees are the following:

- Reflexive: $\forall i \in t\ \ i \preccurlyeq_T i$
- Antisymetric: $\forall i, j \in t\ (i \preccurlyeq_T j \wedge j \preccurlyeq_T i) \rightarrow i = j$
- Transitive: $\forall i, j, k \in t\ (i \preccurlyeq_T j \wedge j \preccurlyeq_T k) \rightarrow i \preccurlyeq_T k$
- Well-founded: $\nexists k \in t\ \forall i \in t\ < i, k > \notin \preccurlyeq_T$

Since trees are binary relations, we denote trees as a set of relationships in the form $T=\{<item_1, item_2>, ..., <item_{n-1}, item_n>\}$, e.g., $T=\{<root, p2>, <p2, a1>, <p2, a2>, <root, p3>, <p3, a2>, <p3, a3>\}$. We can also shorten this notation by representing pairs having the first element in common in the following way: $T=\{<root,\{<p2, \{a1, a2\}>, <p3, \{a2, a3\}>\}>\}$.

*Definition 4.* Let $T = \,< t, \preccurlyeq_T >$ be a tree. Let $<item_i, item_j>$ be an edge in $\preccurlyeq_T$, the parent of $item_j$ is denoted as p($item_j$) = $item_i$. The children of an item are defined by:

$$c(item_i) = \{item_j | < item_i, item_j > \in \preccurlyeq_T\}$$

*Definition 5.* The set of leaves of $T$ is defined as the set of all elements $x \in t$ that has no child items:

$$lf(T) = \{x | c(x) = \emptyset\}$$

*Definition 6.* Let $T = \,< t, \preccurlyeq_T >$ be a relation tree and $i \in t$ be an exploration item. The predecessors of $i$ is defined as:

$$pred(i) = \{y | y \preccurlyeq_T i\}$$

Definition 7. The height of an item $i$ in a tree is the number of predecessors it has:

$$h(i) = |pred(i)|$$

*Definition 8.* The nth level of a tree is defined by the set of items having the height n:

$$L_n(T) = \{x \in t | h(x) = n\}$$

*Definition 9.* each relation $r \in R$, in the dataset <I, R>, is represented as a tree having the relation identifier as its root. We refer to relation sets by specifying their ID preceded by ":". As an example, Fig. 3 shows the relation between publications and their respective authors ($p_i$ denote papes, $a_i$ denote authors).

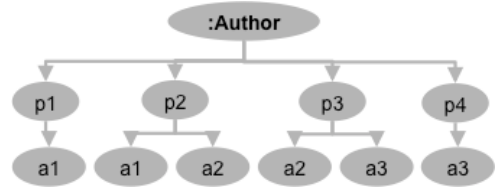

**Fig. 3. Relation tree representing publication-author relationships**

Using the tree notation, the author relation can also be denoted as follows: *Author = {<:Author, {<p2, {a1, a2}>, <p3, {a2, a3}>}>}*. For conciseness purposes, we sometimes omit the root and denote relations in the following way: *:Author = {<p2, {a1, a2}>, <p3, {a2, a3}>}*. We define domain and image properties of relations in sequence.

*Definition 10.* The domain of a relation $Rt$ is the set of all items that are children of the root item:

$$domain(Rt) = \{i | i \in c(root(Rt))\}$$

*Definition 11.* The image of a relation $Rt$ is the set of all second items of the relationships having the domain items in the first position:

$$image(Rt) = \{j | i \in domain(Rt) \land < i, j > \in \preccurlyeq_{Rt}\}$$

As an example of the domain and image properties, let *:Author = {<p1, {a1, a2}>, <p2, {a2, a3}>}* be a relation between publications and authors. The domain and image sets of *:Author* are:

- domain(:Author) = {p1, p2}
- image(:Author) = {a1, a2, a3}

*Definition 12.* The restricted image of a relation $Rt$ on a domain item *domItem*, denoted as *:Rt[domItem]*, is composed of all items of the image of $Rt$ that have *domItem* as its domain:

$$:Rt[domItem] = \{j | < domItem, j > \in \preccurlyeq_{Rt}\}$$

*Definition 13.* The restricted domain of a relation $Rt$ on an image item *imgItem*, denoted as *:Rt⁻¹[imgItem]*, is defined as the set of all items related to a specific image item in $Rt$:

$$:Rt^{-1}[imgItem] = \{j | < j, imgItem > \in \preccurlyeq_{Rt}\}$$

Consider the following examples of restricted image and domain, respectively:

- :Rt[p1] = {a1, a2}
- :Rt⁻¹[a2] = {p1, p2}

*Definition 14.* The join between two relations – *RJoin* - is defined by joining the image set of the former with the domain set of the later, as follows:

$$RJoin(R1, R2) = \{< i, j > | < i, k > \in \preccurlyeq_{R1} \land \exists k < k, j > \in \preccurlyeq_{R2}\}$$

As an example, let *R1 = {<p1, a1>, <p2, a2>}* and *R2 = {<a1, f1>, <a2, f2>}*, the join between these two relations is defined by: *RJoin(R1, R2) = {< p1, f1 >, < p2, f2 >}*. In order to join more than two relations, we make join compositions. Suppose a third relation *R3 = {<f1, i1>, <f2, i2>}*. The join between *R1*, *R2*, and *R3* is: *RJoin(RJoin(R1, R2), R3) = {< p1, i1 >, < p2, i2 >}*.

*Definition 15.* A relation path $R_1,..., R_n$ is an ordered set of relations with non-empty join:

$$RJoin(... (RJoin(R_1, R_2) ...), R_n) \neq \emptyset$$

We denote relation paths by concatenating their identifiers. For example, suppose a relation *:Author = {<p1, a1>, <p2, a2>}* and a relation *:Affiliation = {<a1, f1>, <a2, f2>}*. We denote the path formed by these two relations as: *:Author:Affiliation*

*Definition 16.* A *path set* of an exploration set $S = < s, \preccurlyeq_s >$ is the set of predecessor sets of the leaves of $S$:

$$paths(S) = \bigcup_{l \in lf(s)} \{pred(l) \cup \{l\}\}$$

As an example, consider the set S = {<ids, <a1, {<p1, {f1, f2}>}>>}, the paths of S are:  paths(S) = {{ids, a1, p1, f1}, {ids, a1, p1, f2}}. Each path in *paths(S)* is a sequence in the form $\{a_i\}_{i=1}^{h(l)}$ ordered by the relation $\preccurlyeq_S$, where, *h(l)* is the height of the leaf item *l*, which is the greatest element of the path.

*Definition 17.* Let $p = \{idp, a1, p1, f1\}$ be a path of a set, we define the head of $p$ as the item directly connected to the root of the tree: $head(p) = a1$. The tail of $p$ is the item that has no child: $tail(p) = f1$.

*Definition 18.* The root replacement function $r(Ph, nroot) = Ph'$ maps a path set $Ph$ onto another path set $Ph'$ having the root element (least element) replaced by $nroot$. This function is used for copying a path set from one set to another. Consider the following definition:

$$r(Ph, nroot) = \{< item_i, item_j > \in Ph | \forall < p_i, item_j > \in Ph \ (p_i = root(Ph) \rightarrow item_i = nroot) \wedge (p_i \neq root(P) \rightarrow item_i = p_i)$$

## 2.4 The Exploration Operations

In this section we present formal abstract descriptions of both the exploration process and each exploration operator surveyed from the selected exploration systems.

The exploration process is normally approached as a set of interdependent states [20,37], where each state consists of two components: *Intention* and *Extension*. The *Intention* is a description in some language of the desired set of items in the state. The *Extension* is the actual set of items corresponding to the intention. As an example, consider a keyword search for items matching the keywords "Semantic Web". The intention is the keyword expression and the extension is the set of matched items. In this work, we define an exploration state as an *invocation* of an operator, containing the operator identifier and the actual values bond to each parameter. The extension is the results of the invocation.

*Definition 19.* The exploration process $Po = <St, Dep>$ is a graph where the nodes are exploration states in the set $St$, and the edges are state dependency relations $Dep: St \times St$, such that, each pair of states $< s1, s2 > \in Dep$ represents a relation between a state and its subsequent state. The application of an operation to $s1$ leads the explorer to the state $s2$. Each state $sx \in St$ is an invocation of an exploration operator, which we present in Def. x.

*Definition 20.* The *invocation* of an exploration operator is defined by the operation name and the parameter attributions, represented as a n-tuple of arguments in the form: $<OperationId, Arg_1,..., Arg_n>$.

As an example, consider an initial state $S_0$ whose evaluation generates an exploration set of publications $P = \{<sp, \{p1, p2, p3\}>\}$ and the relations *:Author* and *:Affiliation* that respectively relates publications with authors and authors with their affiliations. The state generated when the user pivots from $S_0$ to the set of authors' affiliations is expressed as follows: $<Pivot, S_0, :Author:Affiliation> \equiv S_0.Pivot(:Author:Affiliation)$

In the pivot invocation above, we define the input exploration set by the identifier of the state whose evaluation generates it. Therefore, instead of using *P*, we use $S_0$, which evaluates to the set *P*. The intentional description of the state is the tuple *<:Pivot, $S_0$, :Author:Affiliation>*, where *:Pivot* is the identifier of the operation, $S_0$ is the input state parameter, and *:Author:Affiliation* is the argument for the *Relation* parameter of the *Pivot* operation. Since each state is generated by the invocation of an exploration operation, we can also describe the exploration process as a functional composition of the operations. Let *Opr = {op₁, op₂, ..., opₙ}* be a set of exploration operations. The exploration process *Po* is described by:

*Po = opₙ(opₙ₋₁(...(op₁(args₁), args₂), argsₙ₋₁), argsₙ),* where *args* is a list of arguments specific for each operation.

In order to aid in the understanding of the formal definitions, we define a generic dataset to demonstrate the semantics of each operation. Consider the following set descriptions:

- A set *T = {<st, {p1, p2, p3, p4}>}* of scientific publications;

- A set *A = {<sa,{a1, a2, a3}>}* of authors;
- A set *F = {<sf, {f1,f2,f3}>}* of authors' affiliations;
- A relation :Author = {<p1,a1>, <p2,a1>, <p3,a2>, <p2,a2>, <p3,a3>, <p4,a3>} between publications and their authors;
- A relation *:Affiliation = {<a1,f1>, <a2,f1>, <a3,f2>}* between authors and affiliations.

The generic dataset $D$ is defined as:

$D = <I, R>$, such that, $I = T \cup A \cup F$ and $R = :Author \cup :Affiliation$

*Definition 21. Pivot(A, Relations): $R \times R^n \rightarrow R$* maps the leaf items of the input exploration set onto another set of related items. Let *rs* be the root item of the result set of a pivot operation. We define *Pivot* as follows:

$$Pivot(A, Relations) = \bigcup_{imgItem \in Relations[lf(A)]} < rs, imgItem >$$

Consider the following examples of *Pivot*:

- *<rs, {a1, a2, a3}>} ← T.Pivot(:Author).* Pivots from the set of publications to the set of related authors;
- *{<rs, {f1, f2, f3}>} ← T.Pivot(:Author:Affiliation).* Pivots from the set of publications to the set of authors' affiliations.

*Definition 22. Refine(A, PathPattern): $R \times (C \times C) \rightarrow R$* filters an exploration set using filtering patterns built upon filters in the set $C$, where $C = \{fp1, fp2, ..., fpn\}$ is the set of filtering predicates available in the environment. A filtering predicate *fp: I→ {true, false}* is defined as a Boolean function that indicates whether an item must be filtered or not. We define a path pattern as follows.

*Definition 23.* Let $S = < s, \leqslant_s >$ be an exploration set. A *path pattern Pa = <F, E>* is a set of filters $F \subseteq C$ and a set of edges between filters, such that, for each $< u, v > \in E$ there is a mapping $< \sigma(u), \sigma(v) > \leqslant_s$. The mapping $\sigma(filter): C \rightarrow I$ matches the nodes in the path having the same height of the nodes in $E$:

$$\sigma = \{< u, j >| u \in F \wedge j \in s \wedge h(u) = h(j)\}$$

Let *True* be a special filter that always evaluates to *true*. The root of the path pattern is always *True*, therefore, for every pattern, *<True, v>* is an edge in $E$ and there is a mapping *<True, root(S)> ∈ $\sigma$*. Fig. 4 shows an exploration set and an example of a path pattern of filters.

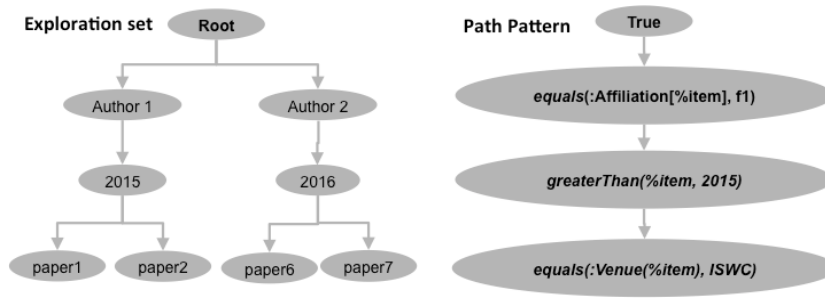

**Fig. 4. An exploration set and a path pattern containing the filters for each level.**

The path pattern of Fig. 4 matches each level of the set, where, authors are filtered by the *:Affiliation* relation, the years are filtered by the predicate *greaterThan(%item, 2015)*, and the papers are filtered by venue. This definition is an adaptation of the graph patterns presented in (AGGARWAL; WANG, 2010, ch. 3 ). Having defined the structure of the path patterns, we can now define the *Refine* operation:

$$Refine(A, PathPattern) = \{path \in paths(A)| \forall < f1, f2 > \in PathPattern$$
$$\exists < \sigma(f1), \sigma(f2) > \in path \wedge f1(\sigma(f1)) \wedge f2(\sigma(f2))\}$$

The *Refine* definition filters every path in the exploration set, such that, for all pairs of filters *<f1, f2>* in the path pattern, there is a matching edge in the path and the filters *f1* and *f2* hold for the matched nodes.

The expressivity of the refine operation is strongly related to the filtering functions and compositions allowed by the tool. The following functions are usually found in exploration tools:

- *equals(item, value2)*: *I×I→{true,false}*: tests the equality of an exploration item against a specific value;
- *equals(item, Relation, value)*: *I×R×I→{true,false}:* tests whether the image of a relation, restricted on an item, is equal to a value;
- *matchAll(item, keywordPattern)*: *I×L^n→{true,false}:* tests if an exploration item matches all keywords in a keyword pattern. Keyword patterns can be represented as a *n*-tuple of literals in a set *L*;
- *matchOne(item, keywordPattern)*: *I×L^n→{true,false}* tests if an exploration item matches at least one of the keywords in a keyword pattern;
- *not(f)*: *{true,false} →{true,false}* negates another filtering function.

Consider the following examples of the *Refine* operation:

- {<rs, {p1, p2}> ← T.Refine(<True, equals(:Author, a1 )>). Finds papers authored by a1;
- {<rs, {p3 , p4}> ← T.Refine(<True, equals(:Author:Affiliation, f2)>). Finds papers authored by authors affiliated to f2;
- {<rs, p3> ← Intersect(
T.Refine(<True, equals(:Author, a2 )>),
T.Refine(<True, equals(:Author, a3)>)
*)*. Finds papers authored by *a2* and *a3* using the *Intersect* operator;
- {<rs, p2, p3, p4> ← Unite(
T.Refine(<True, equals(:Author, a2 )>),
T.Refine(<True, equals(:Author, a3)>)
*)* . Finds papers authored either by *a2* or *a3* using the *Unite* operator.

*Definition 24. Group(A, gR): R × Gf → R* groups an exploration set based on a grouping relation gR ∈ *Gf*, which defines the groups of each item. We call *Gf*: *I × I* the set of all relations available in the environment that can be passed as argument to the *Group* operation. We define *Group* as follows:

$$Group(A, gR) = \bigcup_{path \in paths(A)} \bigcup_{i \in gR[tail(\text{path})]} r(path \cup \{< p(tail(path)), i >, < i, tail(path) >\}$$

$$- < p(tail(path)), tail(path) >, rs)$$

In the formal definition of *Group*, the grouping item $i \in gR[tail(path)]$ adds another level to the input set tree, where the parent of the tail is related to the item *i* and *i* relates to the tail item of the path. However, the tail item will have two parents after this operation, which breaks the condition of a tree. For this reason, we remove the relation between the tail and its previous parent in the last part of the equation using the set difference operation denoted as "−". Consider the following examples of the *Group* operator:

- {<rs,{<a1,{p1,p2}>,<a2,{p2,p3}>,<a3,{p3,p4}>}>}←T.Group(:Author). Groups T by author;
- {<rs,{<f1,{<a1,{p1,p2}>,<a2,p3>}>,<f2,<a3,{p3,p4}>>}>} ← T.Group(:Author).Group(:Affiliation). Groups by author, having the results grouped by affiliation.

*Definition 25. Rank(A, lv, scr): R×N×Sf→R×R* ranks the paths of the input set given a score function *scr* ∈ *Sf* applied to the items in the level *lv* of the input tree *A*, where, *Sf* is the set of available score functions. Let *rs* be the root item of the result set, and *scr(item): I→N* be a score function available in the environment. Let *lv* be the level of the tree containing the items to be scored. We define rank in the following equation:

$$Rank(A, lv, scr) = \{< r(path_i, rs), r(path_j, rs) > | \forall path_i, path_j \in paths(A)$$

$$< path_i, path_j > \leftrightarrow scr(path_i[lv]) \geq scr(path_j[lv])\}$$

For the following examples of the Rank operator, let  *:Year={<p2,2001>,<p1,2002>,<p3,2003>,<p4,2004>}* be a relation between publications and their respective publication years. Let *G = {<sg,{<a1,{p2, p1}>,<a2, {p3, p4}>}>}* be a group of publications by author:

- *{<rs, {p4, p3, p1, p2}>} ← T.Rank(2, :Year[%item])*. Ranks the publications (level 2 of the tree) by year;
- *{<rs, {p2, p1, p3, p4}>} ← T.Rank(2, :Year[%item] * -1)*. Ranks the publications (level 2 of the tree) by year in ascending order;
- {<rs,{<a1, {p1, p2}>,<a2, {p4, p3}>}>} ← G.Rank(3, :Year[%item]). Ranks the third level (leaves) by publication year.

*Definition 26. Correlate(A, B): R × R →R* finds all intermediary pairs of items connecting all source items to all target items. Each path from each source to each target item (many-to-many) is a distinct path in the result tree. Let *A* and *B* be two exploration sets, *rs* be the root item of the result set, and *n* to stand for an arbitrary path length. Let *Prd=lf(A)×lf(B)* be the Cartesian product of the leaf items of *A* and *B*. The set of paths connecting each pair *<item1, item2> ∈ Prd* is defined by:

$$Correlate(A, B) = \{< rs, \{< i_1, i_2 >, ..., < i_{n-1}, i_n > \} > | \; \forall i_j, i_{j+1} < i_j, i_{j+1} > \in R \; \wedge \; < i_1, i_n > \in Prd \; \wedge$$
$$1 \leq j \leq n-1\}$$

The result set of the *Correlate* operation is a set of paths from each origin to each target item, where, origins are the children set of the root and the targets are the leaves. Consider the following examples over the generic dataset:

- *{<rs,<p1, <a1, f1>>>} ← Correlate({p1}, {f1})*. Finds paths between the publication *p1* and the affiliation *f1*
- *{<rs, <p2, {<a1, f1>, <a2, f1>}>>} ← Correlate({p2}, {f1})*. Finds paths between *p2* and *f1*, where, the first path connects through *a1* (*p2→ a1→ f1*) and second path connects through *a2* (*p2→ a2→ f1*).

Here we simplify the definition of the correlation operation in order to avoid excessive details on the operations model and keep it abstract. However, the correlation operation can be specialized with at least two additional patterns. The first parameter is the maximum distance between the origin and the target [5] [23]. The second parameter is a path pattern, which matches each path. A language for pattern definition can be found in [33]. A specialized signature for the *Correlate* operation is as follows:

*Defintion 27. Correlate(A, B, Pattern, maxLengh)* is a specialization of the *Correlate* operator, where, *Pattern* is a path filtering pattern and *maxLength* is the maximum size of the paths to be returned.

Some compositions are very common among path finders. When combined with the *Refine* operation, the user can apply filters that leverage the analysis of connection patterns between the exploration items. For example, suppose a set of paths that connects two hypothetical politicians *pol1* and *pol2*. The user could apply a filtering pattern to keep only paths that contain at least one intermediary node of type "Company". Let *a(%item)* be the set of all ancestors. We can express this task in the following composition:

*Refine(Correlate({pol1}, {pol2}), contains(:Type[a(%item)], :Company))*

The composition above filters all paths having at least one ancestor of the tail items typed as *:Company*. We can also filter the paths using more complex path patterns. For example, we express the task of filtering paths that comprise donations of Companies as follows:

*Refine(Correlate({pol1}, {pol2}), <equals(pol1),*

*    equals(:Type, :Person), equals(:Type, :Company),*

*    equals(:Type, :Donation), equals(pol2)>)*, where, the *Refine* operation filters paths where the politician *pol1* is eventually associated to a person who owns a company that donates to the politician *pol2*.

*Definition 27. HMap(A, lv, f): R×N×Mf → R* applies the mapping function horizontally, i.e., to the children set of each item in the level specified by the *lv* parameter, denoted as *L_{lv}(A)*, as Fig. 5 shows. We can use *HMap* to calculate aggregated values, such as counts and sums, generate transformation sets, and combinations of items within a specific level.

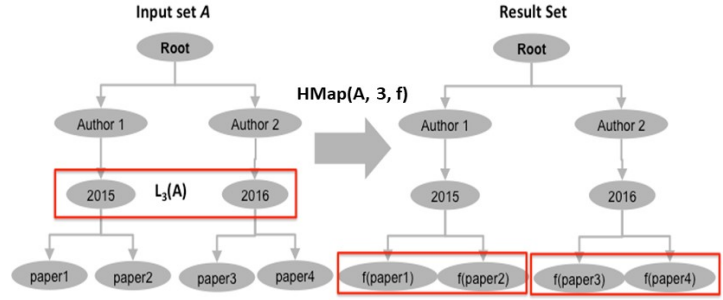

**Fig. 5. An exploration set and a path pattern containing the filters for each level.**

In the mapping of Fig. 5 the level is the next to last (3) and the mapping function *f* is applied to each item of the children sets *c(2015)* and *c(2016)*. The applications of *f* are isolated from each children set. Next, we define the classes of mapping functions that are commonly found in tabular exploration tools as horizontal maps.

*Definition 28. Transformation Mappings – THMap –* are instances of the original higher-order map function usually found in functional programming languages (BIRD; WADLER, 1988 ch. 3), such as Ruby and Python, where, the function *f* is applied to each item of the children set. We define *THMap* as:

$$THMap(A, lv, f) = \{< item, c(item) > | item \in L_{lv}(A) \longrightarrow c(item) =$$
$$\{f(c(item)_{[1]}), f(c(item)_{[2]}), ..., f(c(item)_{[n]})\}\}$$

*THmap* applies the function *f* to each child set of each item in the level *lv* of the input set, represented as $L_{lv}(A)$. *THMap* preserves the structure of the input set, except for the mapping level, which is replaced by the results of *f*. As an example of a transformation mapping, let *M = {<sm, {150.00, 160.50, 135.73}>}* be a set of book prices in *US$*. The user can apply a transformation mapping to get the values in *R$* (Brazilian currency). Suppose the conversion function *rs(value) = value * 3,50*:

$$\{<rs, \{525.00, 561.75, 475.05\}>\} \leftarrow M.THMap(1, rs(\%item))$$

*Definition 29. Aggregation Mappings – AHMap -* reduce the set of children of each item in the level *lv* to a single value, as defined by the following equation:

$$AHMap(A, lv, f) = \{< item, c(item) > | item \in L_{lv}(A) \longrightarrow c(item)$$
$$= \{f(c(item)_{[1]}, f(c(item)_{[2]}, f(... f(c(item)_{[n]}, null) ...)\}\}$$

The *AHMap* operator, applies the function *f* to each child item, where, the next application is the output of the previous application. As an example, for an item having two children, the application of *f* is *f(child1, f(child2, null))*. The *null* value is an initial value to be aggregated with the next applications of *f*. This structure allows, for example, the computation of counts where each application adds one unit to the results of the previous applications, taking the *null* value as 0 (zero). The following example shows how to represent an aggregation mapping in order to count the number of elements of each nesting in the input exploration set. Let *Y = {<sy, {<2005, {p1, p2 ,p3 ,p4}>, <2006, {p5, p6, p7}>}>}* be a set of publications grouped by publication year. In order to map the groups onto counts by year, we do:

*{<rs, {<2005, 4>, <2006, 3>}>} ← Y.AHMap(2, count(%item))*

*Definition 30.* The *Combination Mapping* operator – *CHMap* – applies a n-ary function *f(item₁,...,itemₙ)* to combinations of size *n* of the children set of an item. Let *f* be a combination function and *lv* be the level to map. Let *n* be both the arity of *f* and the size of the combination, and *c(item)ⁿ* be *n* Cartesian products of the children set. We define *CHMap* as:

$$CHMap(A, lv, f) = \{< item, c(item) > | item \in L_{lv}(A) \longrightarrow c(item) = \bigcup_{<i_1, ... i_n> \in c(item)^n} f(i_1, ... i_n) >\}$$

Although the definition above comprises *n* Cartesian products of the children set *c(item)^n*, this function is usually applied to a subset of *c(item)^n*. For example, Tableau and SeCo tools allow the user to map items to values that are combinations of two or more attributes. In these tools, the user can map a set of *Orders* having the columns {*productId, clientId, amount, individualPrice*} to a column *Total = amount*individualPrice*. Considering the children set of each *order* as the total set of values for all columns, the application of the map function is restricted to a subset $Cs \subset c(order)^2$ where the children are values for the attributes *amount* and *individualPrice*.

*Definition 31.* The *Vertical Mapping* operation – *VMap* – applies the mapping function $f$ to each edge of each path of the input set, where, the applications are independent among the paths. Therefore, for any input set $A = <a, \preccurlyeq_A>$, the transformation, aggregation and combination mapping functions are defined respectively as:

$$TVMap(A, f) = \bigcup_{path \in paths(A)} \{\{< m_i, m_j > | < m_i, m_j > = f(< item_i, item_j >) \land < item_i, item_j > \in path\}\}$$

$$AVMap(A, f) = \bigcup_{path \in paths(A)} \{\{< m_i, m_j > | < m_i, m_j > = f(< it_1, it_2 >, f(... (f(< it_{n-1}, it_n >)) \land < it_i, it_j >$$
$$\in path \land 1 \leq i \leq |path| \land 1 \leq j \leq |path|\}\}$$

$$CVMap(A, f) = \bigcup_{path \in paths(A)} \{\{< m_i, m_j > | < m_i, m_j > = f(< it_1, it_2 >, ..., < it_{n-1}, it_n >) \land < it_i, it_j > \in path \land 1$$
$$\leq i \leq |path| \land 1 \leq j \leq |path|\}\}$$

In order to exemplify the application of the *VMap* operation, we take examples of sense-making activities carried out over the results of the *Correlate* operator. For example, the user may want to map each correlation path to a higher level of abstraction, where the relations *<item_i, item_j>* of each correlation path is mapped to a relation between their types. Let *:Type* be a hypothetical relation between items and their types. The following composition expresses this task:

*TVMap(Correlate({pol1},{pol2}), f(%<item1,item2>)=<:Type[item1],:Type[item2]>)*

If we consider the specialized case of correlation that accepts a path pattern – *Correlate(A, B, Pattern, maxLenght)* – The composition of *TVMap* with *Correlate* can be used to generate the abstract path patterns that can be the input of another correlation operation or a query over the dataset. In fact, this composition describes the Fusion tool [5].

*Definition 32. Unite(A, B): $R \times R \rightarrow R$* receives two exploration sets and unites their path sets. Let *rs* be the root item of the result set. We define the union operation as:

$$Unite(A, B) = r(paths(A), rs) \cup r(paths(B), rs)$$

The results of *Unite* is the union between the paths of *A* and *B*, mapped to common root item *rs*.

*Definition 33. Intersect(A, B): $R \times R \rightarrow R$* computes the intersection between the paths of the input sets. Let *rs* be the root item of the result set. We define *Intersect* as:

$$Intersect(A, B) = r(paths(A), rs) \cap r(paths(B), rs)$$

The set of paths of the result set is the intersection between the paths of *A* and *B* with the root replaced by the *rs* item.

*Definition 34. Diff(A, B): $R \times R \rightarrow R$* computes the difference between the paths of *A* and *B*. let *rs* be the root item of the result set. We define *Diff* as follows:

$$Diff(A, B) = r(paths(A), rs) - r(paths(B), rs)$$

The paths in the result set are all paths in *A* that do not appear in *B* under the same root *rs*. In order to exemplify the application of the set operations *Unite*, *Intersect*, and *Diff*, let A = {*<sa, {p1, p2, p3}>*} and B ={*<sb, {p2, p3, p5}>*} be two sets of hypothetical publications. Let C = {*<sc, {<a1, {p1, p2, p3}>, <a2, {p3, p4}>}>*} and D = {*<sd, {<a1, {p2, p3, p5}>, <a2, {p3, p5, p6}>,<a3, p8>}>*} be two grouped sets:

- {<rs,{ p1, p2, p3, p4, p5}>} ← Unite(A, B)
- {<rs, {<a1, {p1, p2, p3}>, <a2, {p3, p4, p5, p6 }>, <a3, {p8, p9}>}>} ← Unite(C, D)
- {<rs, {p2, p3}>} ← Intersect (A, B)
- {<rs, {<a1, {p2, p3}>, <a2, p3>}>} ← Intersect (C, D)
- {<rs, p1>} ← Diff(A, B)
- {<rs,{<a1, p1>,<a2, p4>}>} ← Diff(C, D)

## 3   CASE STUDY: EVALUATING A SCIENTIFIC PAPER

Here we present a case study devised to evaluate the framework of exploration operations when used to describe exploration strategies in real and well-documented problematic situations. The main goals of the case studies are: 1 – demonstrate the expressive power of the framework to describe complex task solutions in terms of sequence applications of the exploration operations; 2 – demonstrate the usage of the framework as an epistemic tool for devising alternative sequences of steps; 3 – demonstrate possible reuse and adaptation scenarios for explorations.

We selected the case study in the scientific publications field to demonstrate the operations. We also selected the Open Citations [31] dataset for the simulation, which is an RDF dataset of scientific publications. The following task was presented in [26]:

Consider a reviewer evaluating a scientific paper. In order to do so, the user can take the following strategy:

1.    Analyze the age of the citations: the reviewer extracts the years of each citation and calculates, for example, the mean year;

2.    Check the lack of citations to relevant publications: the reviewer can extract the keywords of the paper and issue a keyword search for related papers; Rank the articles by the number of incoming citations; Keep the first 20 articles; Differentiate the two sets and verify which ones are not in the bibliography of the paper;

3.    Analyze the degree of "self-citations": the reviewer analyzes how self-referential is the paper. A self-citation can be either a citation of previous works of one of the authors or citations from authors of the same research group;

4.    Evaluate if the paper fits to the scope of a venue: the reviewer might count the number of citations published in the same venue as an indicator of how adequate the paper is to the targeted venue.

Fig. 6 shows a representation of the Open Citations schema slightly adapted for demonstrations purposes.

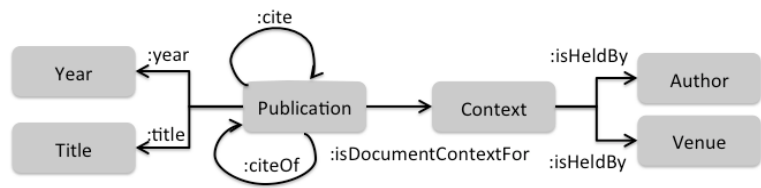

**Fig. 6. Open citations summarized schema describing relationships between publications, years, authors, and publication venues**

We represent the exploration strategy described in the problem statement as the following sequence of steps in our framework. Let $D$ be a dataset of papers, and $p$ be a unitary set having the paper under reviewing:

1.    $S_1 \leftarrow$ p.$Pivot$(:cite)

2.    $S_2 \leftarrow S_1.Pivot$(:year)

3.    $S_3 \leftarrow S_2.AHMap$($mean$)

4.    $S_4 \leftarrow$ D.$Refine$(matchAll("Semantic Web"))

5.    $S_5 \leftarrow S_4.Group$(:cite)

6.    $S_6 \leftarrow S_5.AHMap$(2, count)

7. $S_7 \leftarrow S_6.Rank(1, c(\%item))[0..19]$
8. $S_8 \leftarrow S_7.Diff(S_1)$

9. $S_9 \leftarrow p.Pivot(:isContextFor:isHeldBy)$

10. $S_{10} \leftarrow S_9.Refine(equals(:type, Author))$
11. $S_{11} \leftarrow S_{10}.Pivot(:isHeldByOf:isContextForOf)$
12. $S_{12} \leftarrow S_1.Intersect(S_{11})$
13. $S_{13} \leftarrow S_{12}.AHMap(count)$

14. $S_{14} \leftarrow p.Pivot(:isContextFor:isHeldBy)$
15. $S_{15} \leftarrow S_{14}.Refine(equals(:type, Venue))$
16. $S_{16} \leftarrow S_1.Refine(equals(:isContextFor:isHeldBy, S_{15}))$
17. $S_{17} \leftarrow S_{16}.AHMap(count)$

Step 1 pivots from the set containing the paper under review to the set of citations through the relation *:cite*. Steps 2 and 3 pivot to the citations' years of publication and computes the average year, respectively.

In order to find papers that are relevant to the field but were not cited, in step 4 the user filters the dataset to find papers related to the Semantic Web area using a keyword filter. In steps 5 and 6, the user, first groups semantic web papers by their outgoing citations using the relation *:cite*, and then counts the groups using an aggregation mapping. In the succeeding step, the user ranks the semantic web papers by their incoming citations count, to measure their relative relevance, and keeps only the first twenty. A set difference is carried out in step 8 to find the relevant papers that were not in the set of citations of the paper being reviewed.

Steps 9 to 13 aim at verifying how self-referential the paper is. In order to do so, the user tries to reach the set of papers published by the authors by pivoting to the publication holders (step 9) and obtaining the authors of the reviewed paper (step 10). Then the user pivots back from the authors to the authors' publications (steps 11). The path *:isContextFor:isHeldBy* relates papers and their holders (authors or venues) in the Open Citations dataset. Subsequently, the user counts the intersections with the set of citations (steps 12 and 13). Finally, the user calculates how many citations were published in the same journal as a measure of adequacy of the reviewed paper to the submitted journal (steps 14 to 17).

## 3.1 Alternative Strategies

As an example of alternative strategies, we can consider the steps related to the analysis of how self-referential the paper *p* is (steps 9 to 13). The strategy used is checking how many of the citations of *p* were authored by the authors themselves. From the step 11, the user could extend this task to compare the citations against the papers published by the same group of researchers. Since there is no schema relation for research groups, the explorer tries to approximate them by adding the papers published by the co-authors of the authors of *p* in the comparison. The original steps 12 and 13 can be replaced by the following sequence:

$S_{12}' \leftarrow S_{11}.Pivot(:isContextFor:isHeldBy)$
$S_{13}' \leftarrow S_{12}'.Refine(equals(:type, Author))$
$S_{14}' \leftarrow D.Refine(and(equals(:type, Publication), equalsOne(:isDocumentContextFor:isHeldBy, S_{13}'))$
$S_{15}' \leftarrow S_1.Intersect(S_{14}')$
$S_{16}' \leftarrow S_{15}'.Map(count)$

In the alternative step 12, the user pivots from the publications of the authors of *p* (set $S_{11}$) to the set of all holders. Next in step 13, the user refines the set of holders in order to keep only those that are authors, thus, excluding the venues. The set $S_{13}'$ includes both the authors of *p* and their co-authors. In the alternative step 14, the user applies a conjunctive filter for all items of the type *Publication* having at least one author in the set $S_{13}'$ (*equalsOne* filter). In step 15, the user computes the intersection between the citations in $S_1$ with the set of publications of both the authors of $p$ and their co-authors. The final step is to count the intersection results, which can be found in $S_{16}'$.

Another interesting case of alternative strategy concerns the steps 16 and 17 of the original task. This step aims at refining the citations published in the same venue as $p$. Some users more familiarized with browsing actions may prefer to navigate the set of venues of the citations, in order to learn and make sense of them, before refining. Therefore, the following sequence of steps may replace the original steps 16 and 17:

$S_{16}' \leftarrow S_1.Pivot(:$isDocumentContextFor:isHeldBy$)$

$S_{17}' \leftarrow S_{16}'.Refine($equals$(:$type, *Venue*$))$

$\quad S_{18} \leftarrow S_{17}'.Pivot($isHeldByOf:isDocumentContextForOf$)$

$\quad S_{19} \leftarrow S_1.$Intersect$(S_{18})$

$\quad S_{20} \leftarrow S_{29}.Map(count)$

In the alternative steps 16 and 17, the user decided to pivot from the citations to the set of holders and refines the venues in order to make sense of them. Next, the user pivots back to the set of all publications held by the venues of the citations in step 18, and intersect with the citations. The last step is to count the amount of intersections in step 20.

Although further analyzes should be carried out, we believe that the sequence of steps employed may reveal some characteristics of the user profile. For example, the alternative sequence to count citations published in the same venue as the reviewing paper may reveal a user that has more familiarity or prefer browsing operations. Another possible case is the lack of knowledge concerning publications venues of a specific area, which forced him/her to explore this area of the dataset. We are not judging here which sequence is better, since there are many variables involved. The main goal, though, is to emphasize the utility of the framework as an epistemic tool for investigating alternative exploration strategies.

### 3.2 Reuse Scenario

As a reuse case, we identified a scenario of an inexperienced reviewer carrying out the same task, s/he can miss the self-citation analysis step. In this case, at least two actions can be taken. 1 – the system can identify in its database that other users have carried out the same task and their solutions have a greater number of sessions and steps, thus, the system suggests additional actions; 2 – before finishing the task, the user issues a query for existing previous compositions for the same task and verifies that other reviewers have considered further steps. The user, therefore, decides to incorporate the self-citation analysis to his solution and compute the exploration graph.

The main goal of the reuse in this case is the transference of knowledge not only with regards to the results of the tasks but also concerning the resolution processes. Therefore, new users can draw upon the experience of previous users to aid their task resolution strategies.

## 4   EVALUATION FRAMEWORK

The evaluation framework addresses two main aspects of the exploration. The tactical aspect, which describes the most common atomic operations, and the strategic aspect that describes composition patterns of tactics to solve complex information problems.

The tactical analysis can help the designer in assessing what a user can accomplish within the scope of a single action. However, it lacks semantics for describing exploration strategies, which requires the description of the possible sequences of actions in time. Therefore, we characterize the possible exploration strategies allowed by the tool using strategy analyzes. We employ the following procedure to derive evaluations and comparisons of exploration tools:

1.   For each Operation in the framework

    a.   Analyze how many are covered through the interface;

b.  For each parameter, analyze the range of arguments that can be passed according the established criteria in the tactical map;

.   List possible specializations

Build a context-free-grammar to analyze the possible sequences of expressions;

Derive assessments and possible improvements.

## 4.1 Tactical Analysis

We describe the tactical aspects in terms of the signatures of each operation, comprising the description of parameters and result sets. Thus, we devised a set of attributes to assess the range of arguments that can be passed as parameters for each operation, based on the formal description of the parameters. We organized the tactical analysis as a decision tree containing questions (rectangles) and possible answers for the operators (ellipses). Fig. 7 shows the decision tree for tactical analysis.

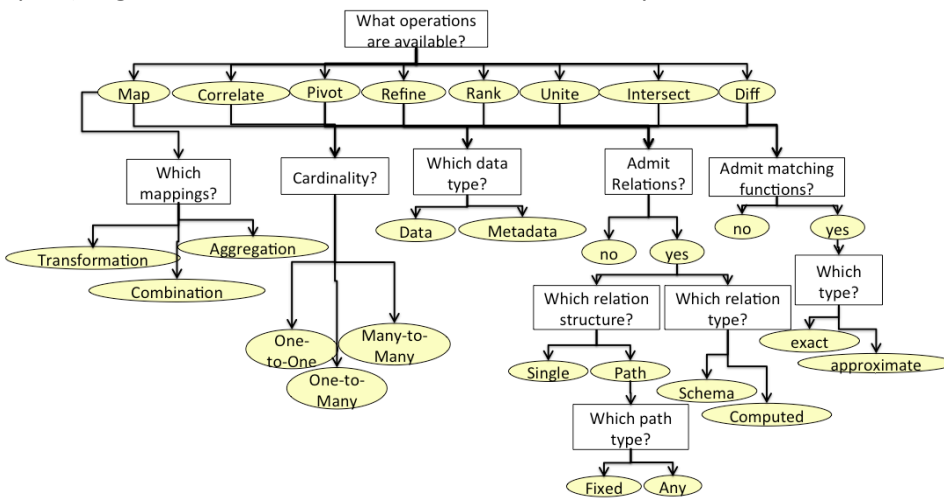

**Fig. 7. Design rationale representation of the central questions and answers for tactical analysis**

The tactical analysis tree allows the assessment of what operations are available through the interface and which arguments they can receive, informing how they can be used in the exploration. The rectangles represent questions for each operation and the ellipses are the possible answers. The root of the tree is the main question "*What operations* are *available?*" and the possible answers are the framework operations. Next, there is a set of questions for each operation. As an example, some operations receive relations as arguments, such as *Pivot* and *Refine*. If the answer for the question "*Admit relations?*" is "*yes*", this leads to questions concerning both the relation structure, which can be a "*single*" relation or a relation "*path*", and the relation type, which can be a relation from the "*schema*" or a "*computed*" relation generated along the exploration process. The descriptions of the analysis questions and answers are as follows:

• "Cardinality?": characterizes whether the operation is a mapping from one item to another item ("One-to-One"), one item to many items ("One-to-Many"), or many items to many items ("Many-to-Many");

• "Which data Type?": specifies whether the operation accepts "data" or "metadata" as input. Our model makes no difference between data and metadata, i.e., both types can be the input of exploration operations. The relevance of this characteristic is due to schema learning. As an example, consider a faceted search interface manipulating a dataset with hundreds of facets. If the facets could be ranked and refined by relevance, it could leverage learning for future steps.

• "*Which relation type?*": characterizes whether the operation accepts *"schema"* relations or *"computed"* relations, or both;

• "*Which relation structure?*": characterizes whether the operation accepts relation *"paths"* or only *"single"* relations. In case of relation paths, they can be *"Any"* or *"Fixed"*, where the former means that the user can use any property path of the dataset, while the latter informs that only fixed and preconfigured paths can be used.

- *"Admit matching functions?"*: the refine operation receives matching predicates for items' comparisons that can be of two types: *"exact"* or *"approximate"*;
- *"Which mappings?"*:  This question characterizes the types of mapping functions available for Map operations. The possible types are *"aggregation"*, which maps a whole set of items onto a single aggregated value, *"combination"*, which combines two or more values, and "transformation", which maps a set of items onto another equivalent set of items that are some kind of transformation of the original items. As examples of the latter, consider date format transformations and currency conversions.

As an example of tactical analysis, we choose gfacet and Search Computing tool (SeCo) and compare their tactical expressivity. Fig .8 presents a gfacet screen with two interrelated sets of items.

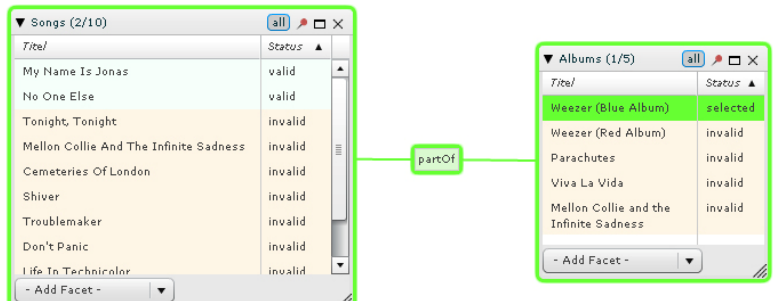

**Fig. 8. gfacet's multifocal scre[...]ms connected by the "partOf"**

In gfacet the user starts with a k[...] the workspace. Then, the user can select a facet, i.e., a relation w[...]t. gfacet adds the new set and a relation to the workspace. Once ha[...]er can select an item from one set and gfacet filters the items that[...]n Fig. 8 the user filters all items that are part of the album "Weeze[...]s the ability to filter sets along pahts with two or more relations. I[...]

The SeCo tool is an exploration t[...]ects make SeCo an interesting case of analysis. First, its design wa[...] which is a well-known episodic model for describing exploration b[...]cond, SeCo presents a formal model of exploration operations, r[...]e SeCoQL. Fig. 9 shows screenshots of SeCo.

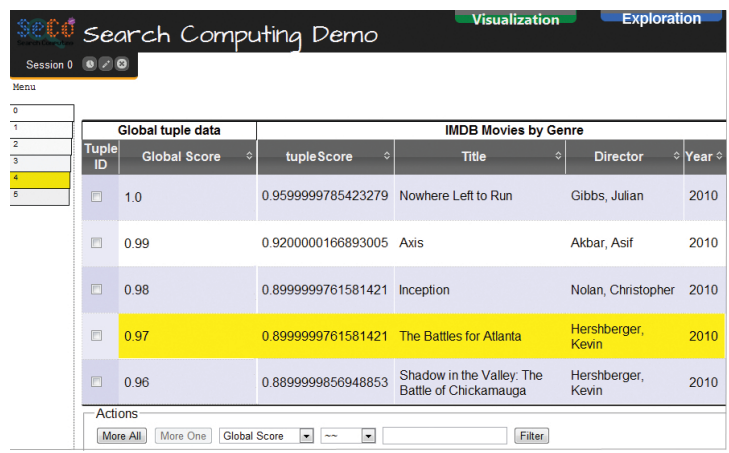

**Fig. 9. Search Computing's unifocal screen presenting a set of movies.**

In SeCo, the user can start the task by applying filters on items types and their attributes. SeCo presents the results as a table that can be further processed. The user can also add related items by selecting relations through the expansion mechanism and joining the rows in the current table view. The rows can also be ranked and grouped.

Using the tactical map approach, we can both analyze the available operations of gfacet and SeCo and draw expressivity comparisons within the scope of single actions. Fig. 10 and Fig. 11 presents the tactical maps of gfacet and search computing, respectively.

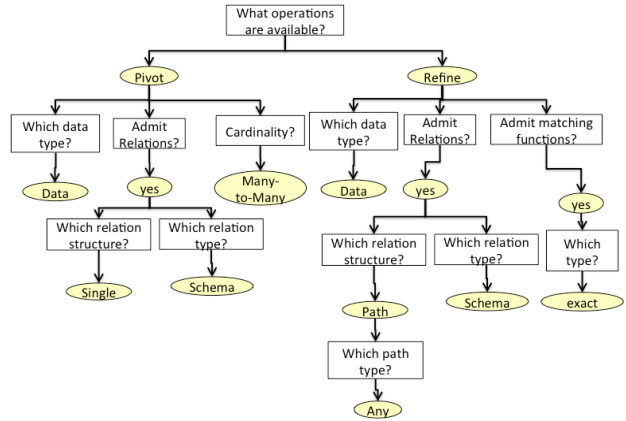

**Fig. 10. gfacet's tactical map describing the expressivity of each atomic action.**

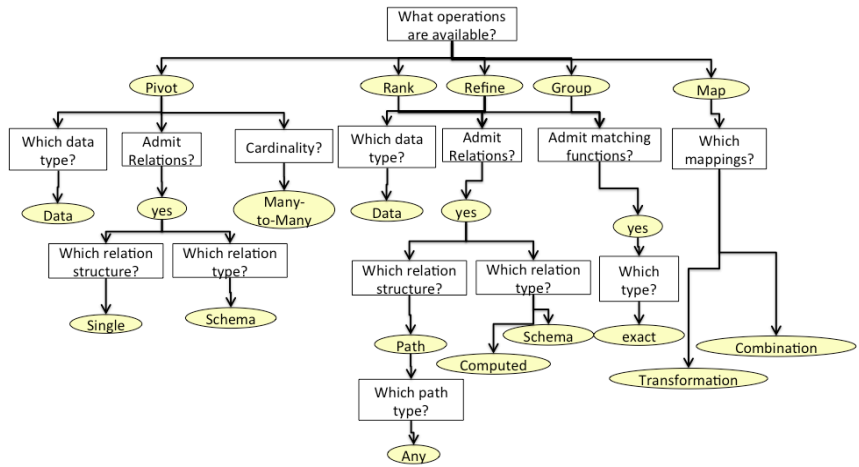

**Fig. 11. Search computing's tactical map describing the expressivity of each atomic action.**

From the gfacet's tactical map (Fig. 10) we can draw the following conclusions:

• Only two operators are available: Pivot and Refine;

• Pivot's cardinality is "Many-to-Many", hence, it is possible to navigate from multiple items to multiple related items;

• Since the relation structure of the Pivot is "single", there is no possibility of pivoting through relation paths;

• Since the relation structure of the refine operation is a "path", it is possible to filter items separated by two or more relations, i.e., indirectly related;

• It is not possible to refine using similarity measures, such as string distance, since the value of the match type attribute is "exact", for the Refine operation;

• gfacet does not allow manipulation of metadata, which may hinder schema learning.

By assessing the tactical map of SeCo, presented in Fig. 11, we can draw the following comparisons and conclusions:

- SeCo has a more complete set of operations than gfacet: Pivot, Refine, Group, Rank, and Map, noting that Map is uncommon among exploration tools.

- Despite the big difference in result set presentation between SeCo and gfacet (tabular vs graph based), the Pivot operation has the same characteristics in both - it does not admit relation paths and are restricted to schema relations.

- There is a difference in the relation type of the Refine operation. In SeCo, Refine accepts not only schema, but also computed relations. This is due to the possibility of creating new columns, not provided by the schema, and using their values in filtering actions.

- SeCo allows mappings to combined fields based on two or more fields, such as "TotalPrice = Theatre.Price + Restaurant.AvgPrice" [10]. However, as far as we could determine, it is not possible to map using aggregation functions.

## 4.2 Strategic Analysis

The strategic analysis step aims at assessing the expressiveness of the tool from the point of view of the range of strategies the user can employ to solve exploration problems. The range of different functional compositions that is allowed by the tool when we abstract interface and interaction details defines the possible strategies. Once the operations framework is defined, we represent the dependencies between the operations using context free grammars, enabling analyses such as which operations can be the input of another. In order to define the grammars of each tool, we first establish the basic notation:

- **Nonterminal symbols:** we use capital letters for non-terminal symbols: S, R, O, etc.

- **Terminal** symbols: terminal symbols are operation names represented in lowercase and Italic: *intersect(), unite(), refine()*. Let "*s0*" be a terminal representing the starting point or a dataset reference.

- **Production rules:** defined as non-terminal-symbol replacements. Consider the following examples and some possible derivations (sentences in the grammar), respectively. Let ":=" stand for "derives":

R → *refine*(R) | s0 :=
refine(refine(s0))
S → *refine*(S) | *pivot*(S) | s0 :=
refine(pivot(s0)),
refine(pivot(pivot(s0)))

- **Operation arguments:** even though we use filters and relations to leverage the understanding of the examples, for the sake of simplicity, we describe the grammars only in terms of the input arguments of the operations. Other types of arguments, such as refine filters or pivot relations are abstracted.

- **Refinements through compositions:** some faceted systems allow the user to pivot multiple times, using schema relations, and apply a restriction at some point. This restriction is back propagated to the previous sets. Let *s0* be a set of publications. The following composition illustrates this case: *refine*(*pivot*(*pivot*(s0, :Author), :BirthPlace), *equals*(:PartOf, "Brazil"))!

The composition above expresses a user navigating from *s0* to the set of authors, and then, to their birthplaces. The birthplaces are restricted to those being a location in Brazil. In order to propagate this restriction to the set of authors, and subsequently, to the initial set of papers, we use the symbol "!" appended to the refine operation.

- **Branching:** some tools, such as gfacet and Parallel Faceted Browser, allow repeated applications of exploration operations over the same set of items, where the results of an application does not interfere in the results of the next, i.e., the applications are independent from each other. Consider the case of a user applying two independent refinements to a hypothetical set of publications *P* in our framework:

  o   $P_1 \leftarrow$ P.Refine(equals(:Author, a1))
  o   $P_2 \leftarrow$ P.Refine(equals(:Author, a2))

Fig. 12 shows a graphical view of the applications.

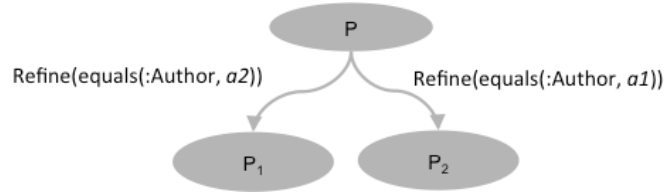

**Fig. 12. Parallel applications of the Refine operator.**

The set *P₁* contains publications of the author *a1* and the set *P₂* contains the publications of the author *a2*. The application of the second refinement is independent of the application of the first refinement since both refinements have the set *P* as input. Moreover, both *P₁* and *P₂* are accessible through the interface for further analyzes. For similar cases, we introduce the *branch* operator as follows:

*Definition 35. branch(InputExp, Exp₁,Exp₂)* receives  an input expression *InputExp* and two expressions *Exp₁* and *Exp₂,* and returns two independent results. The input of both *Exp₁* and *Exp₂* is the result set of *InputExp*. Let the terminal *irs* denotes the result set of the input expression. The refinements of Fig. 12 can be expressed in the following way: *branch(P, refine(irs), refine(irs))*

The branching operator should not be confused with the set operators *unite, intersect,* and *diff*. The set operators receives two input expressions and return a single result set that is a combination of the two inputs.

## 4.3 Comparing Faceted Search Tools

In order to test our evaluation method, we looked at a set of 13 faceted search systems and show how we can categorize them in terms of their strategy grammars. We organize them in increasing order of expressiveness.

For each grammar, we present task examples and an assessment. Consider the following hypothetical dataset for the examples: let *s0* be a dataset of publications having the relations *:Author = {<p1, a1>, <p1, a2>, <p2, a3>}, :Year = {<p1, 2002>, <p2, 2005>, <p3, 2007>}, :Venue = {<p1, ISWC>, <p2, ISWC>, <p3, WWW>}*. We describe the faceted search grammars in the next subsections.

*4.3.1   Faceted Search Version 1. .*  This grammar describes the earliest faceted search tools, such as Flamenco, mspace, and also most of the e-commerce websites: S → refine(S | s0).

The version 1 grammar only features combinations of refine operations, allowing the application of conjunctive filters. A task example leveraged by such grammar is "Find papers authored by author *a1* that were published in ISWC in 2002". However, some task categories are unfeasible for this grammar, such as, disjunctive filtering, comparison tasks, and navigation tasks. Therefore tasks such as "Find papers published either in ISWC or ESWC", "Trace similarities and differences between publication behavior of two researchers", and "Navigate to the authors of the papers and filter their affiliation" are very difficult to accomplish.

*4.3.2   Faceted Search Version 2.*   This grammar adds the possibility of branching a result set and applying independent filters, which leverages comparison tasks:

S → *branch*(s0, S, S) | R

R → refine(R | s0)

The addition of the branch operation is actually the contribution of the Parallel Faceted Browser. Therefore, it is possible to approach some comparison tasks without using external tools. For example, the task of comparing the publication venues of two researchers can be solved more easily with the following expression:

branch(s0,
*refine*(irs, *equals*(:VenueOf:Author, a1)),
*refine*(irs, *equals*(:VenueOf:Author, a2))

)

### 4.3.3    Faceted Search Version 3.

The version 3 grammar aims at adding navigation possibilities to the traditional faceted filtering captured in the version 2. Therefore, the version 2 grammar allows the user to not only restrict the current set of items using conjunctive filters, but also to navigate to related items and change the focus from one type to another, e.g., from publications to their authors. This grammar describes the possible strategies in /facet, tfacet, gfacet, and Rhizomer:

S → P | R | *branch*(S, P, P)

R → refine(R | P | s0)!

P → pivot(R | s0 | P)

The possibility of combining *Pivot* with *Refine* augmented considerably the expressivity of the tools. Now, the user can browse through different data types and relations and apply restrictions to more than one type of item. For example, we can execute the task "Find papers whose authors' affiliations are PUC-Rio" by navigating to the set of authors, recognize and learn their relationships, navigating to the related affiliations, and finally applying the desired restrictions. Let :*Abbr* be the relation between affiliations and their abbreviated names, the following expression represents this solution strategy:

*refine*(

      *pivot*(

            *pivot*(s0, :Author),

            :Affiliation

      ),

      *equals*(:Abbr, "PUC-Rio")

)!

In the task example 2, the user changed the focus from papers to authors, and then to affiliations (*pivot(pivot(s0, :Author), :Affiliation)*). Next, the set of affiliations was filtered to the one having "*PUC-Rio*" as abbreviated name, and the restriction was propagated to the intermediary sets as a result of the operator "!" described in the previous section.

### 4.3.4    Faceted Search Version 4.

This grammar adds the *Unite* operator, which allows the expression of disjunctive filters. Moreover, the results of any operator can be the input of the remaining ones, hence, allowing more filtering possibilities and comparative analyses:

S → *branch*(S, S, S) | O| R |R! | P

O → *intersect*(S, S) | *unite*(S, S)

R → *refine*(O| R | P | s0)

P → *pivot*(O| R |P | s0*)*

Considering faceted search environments, the version 4 grammar is the most expressive, allowing any combinations of the operators and branching. This grammar describes Sewelis and SemFacet tools. The limitations of this grammar concerns the tactical aspects, since operations such as, *Rank*, *Correlate*, *Map*, and *Group,* are missing. An example of a task leveraged by this grammar is the task example 3: "Find papers published in ISWC or ESWC whose authors' affiliations are PUC-Rio or UFRJ":

intersect(

union(

*refine*(s0, *equals*(:Venue, ISWC)),

*refine*(s0, *equals*(:Venue, ESWC))

),

union(

*refine*(s0, *equals*(:Author:Affiliation, PUC-Rio)),

*refine*(s0, *equals*(:Author:Affiliation, UFRJ))

)

)

Table x presents a summary of the strategic analysis of the state-of-the-art faceted search tools.

**Table 1. Exploration grammars of state-of-the-art tools**

| Tools | Grammars |
|---|---|
| Flamenco, Mspace, Faceted Wikipedia Search | S → *refine*(S \| *s0*) |
| Parallel Faceted Browser | S → *branch*(s0, S, S) \| R<br>R → *refine(*R \| *s0)* |
| Humboldt, Parallax | S → P \| R<br>R → *refine(P \| s0)* \| *intersect*(R , R)<br>P → *pivot(R \| s0 \| P)* |
| /facet, gfacet, tfacet, Rhizomer, BrowseRDF | S → P \| R \| *branch*(S, P, P)<br>R → *refine(R \| P \| s0)*!<br>P → *pivot(R \| P \| s0)* |
| Sewellis, SemFacet | S → branch(S, S, S) \| O\| R \|R! \| P<br>O → *intersect*(S, S) \| *unite*(S, S)<br>R → *refine(*O\| R \| P \| s0)*<br>P → *pivot(*O\| R \|P \| s0)* |

## 5 IMPLEMENTATION

In this section, we present details of the application of the concepts of the framework of exploration operations in the design and implementation of a real exploration tool, Xplain[3].

The functional layer contains the exploration operations of the proposed framework accessible through a specific DSL (Domain Specific Language) in Ruby language. Fig. 13 shows the task presented in the case study of section 3, represented in the exploration DSL.

```
1   #Subtask1: mean publication year of the references
2   s1 = p.pivot{relation "cite"}
3   s2 = s1.pivot{relation "year"}
4   s3 = s2.map{mean}
5
6   #Subtask2: find missing relevant citations
7   s4 = d.refine{match_all "Semantic Web"}
8   s5 = s4.group{relation "cite"}
9   s6 = s5.map{count}
10  s7 = s6.rank{score_by_image level: 1}[0..19]
11  s8 = s7.diff s1
12
13  #Subtask3: finding the amount of self-citations of the paper
14  s9 = p.pivot{relation "isDocumentContext", "isHeldBy"}
15  s10 = s9.pivot{relation inverse("isHeldBy"), inverse("isDocumentContextFor")}
16  s11 = s1.intersect s10
17  s12 = s11.map{count}
```

**Fig. 13. DSL representation of the solution strategy for the paper review case study presented in section 3.**

Let *d* be a reference to the whole dataset. The result sets achieved by the script of Fig. 13 are stored in an exploration session and each exploration set keeps a reference to input sets they depend on. Therefore, there is a dependency graph that can be shown in the interface as an exploration trail and further manipulated for reuse purposes.

We designed and implemented the Web interface of the Xplain tool on top of the DSL. The client sends DSL expressions to the server, which executes them, and returns the resulting exploration sets presented in the Xplain screen, as Fig. 14 shows.

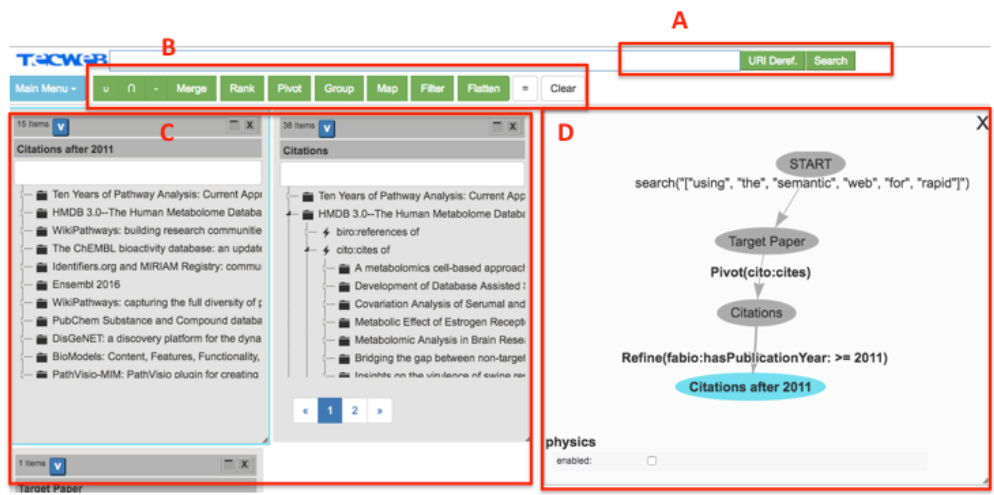

**Fig. 14. The interface of the Xplain environment. (A) keyword search controls; (B) Exploration operations toolbar; (C) Exploration sets area; (D) Exploration trail view**

Fig. 14 shows the main concepts of the proposed framework of exploration operations implemented in Xplain system, named: Exploration Sets (Fig. 14 C); Exploration Items and Relations (the items held by the sets in Fig. 14 C); Exploration Actions (Fig. 14 A and B); The Exploration Trail presenting both the relationships between sets and a history of actions (Fig. 14 D).

## 6   RELATED WORKS

The need for separating visual representations from processing operations has been established in the visualization area presenting taxonomies, typologies, and ontologies addressing at least these two concerns [3,4,12,17,35]. Chi's work [17] divides the design of a visualization system in a sequence of stages in a pipeline that receives raw data as input and generates interactive visualization of the raw data as output. Each *pipeline* stage receives a set of data items, applies processing operations to transform the data, and passes the transformed data to the next stage. "scroll", "zoom", "filter", "rotate", and "scale" are operations that can be carried out in the view resulting from the pipeline. The work in [4] presents a taxonomy of analytic operations for describing visualization tasks containing, for example, "Cluster", "Filter", "Sort", and "Correlate" operations. The work in [12] presents a typology of abstract visualization tasks addressing the *Why*, *How*, and *What* aspects of a visualization task independently of the kind of visualization and of the task domain. The *Why* concerns the goals, such as "discover" new information, "present" data, or "explore". The *How* presents the actions to achieve the goals, such as "select", "navigate", "filter", and "aggregate". The *What* describes input and output resources handled by the tasks. These approaches are valuable to promote some degree of separation between the description of users' goals, tasks, and operations from visual encoding details but their lack of formality makes it hard to analyze where they overlap and what are the differences. Moreover, they do not present detailed discussions of design issues with respect to a given well-defined conceptualization of exploration processes and strategies.

Visualization systems are concerned with encoding data in a visual representation to foster human cognitive perception. Although interactive visualizations can be used to explore a dataset to some extent, supporting exploration *behavior* is not its main goal [38]. Moreover, even in interactive visualizations, the user is usually restricted to a specific visual representation of the data aiming at highlighting a certain set of dimensions. Since exploratory search tasks tend to be general and multifaceted [40], it is very difficult to know in advance which data dimensions will suffice. Therefore, an exploration environment should support a broader class of tasks that may even include sense-making activities and manipulations of the raw data in order to select proper dimensions to encode in visual representations. In this context, one advantage of our reference framework is that it allows designers to situate visualization concerns in the design of exploration environments. Within the framework, the projection of the data onto a visual counterpart along with interaction controls and dialogue structures is a concern of the interaction/interface layer. The data processing interactions can be designed on top of the same framework of operations whose inputs are visually encoded items and relations.

Beyond the works in visualization field, there are some works addressing issues related to the broader exploration field. The work in [10] presents a similar architectural view of exploration environments and also abstracts the functional aspects in the SeCoQL exploration language. However, it presents a restricted set of operators containing only refine, pivot, and ranking, and does not approach interface concerns in detail. In a similar way, the works in [30], [34], [20], [37] present formal descriptions for faceted searches that, as Table 1 shows, are subsets of the exploration operations framework presented in this work.

The works in [2] and [41] propose a separation of interface details from the underlying features, but there, the main goal is to present a common evaluation framework for comparison purposes and not a detailed design space discussion. Moreover, the features are not formally defined, which is the source of many ambiguities, such as different interactions for an operation being mistakenly understood as being distinct features.

# 7  CONCLUSION AND FUTURE WORKS

Although the separation of the conceptual model of user operations from the interface details is a widely accepted principle in HCI design, it has hitherto not been properly applied in the context of information exploration tools. Consequently, it is difficult to assess and compare the adequacy of these tools for information exploration tasks. Given this scenario, this work proposed a novel framework of exploration operations expressive enough to describe at least the majority of state-of-the-art environments and tasks.

In order to build the framework we analyzed more than 20 exploration environments and models published in the literature, such as faceted search tools and models, set-oriented browsers, tabular data processors, and visualization tools and models. From the analyses, we derived a collated list of features that were generalized and formalized as abstract data processing operators.

The framework allows both the formal description of explorations strategies and a more precise assessment of the functional aspects of exploration environments, independently of how the operations are articulated thorough the interface. Therefore, we can compare the extent of the exploration support of different tools using the same set of operations, as we demonstrated in section 4.

The present work establishes a theoretical base for future researches on both the design of exploration environments and user studies. We presented a discussion of the separation of interface and interaction issues from the underlying functional operators. However, the interface design also comprises a layer of abstract widgets that may lead to an abstract interface model. As an example, the selection and definition of operators and parameters must be supported by interactive widgets that should express exploration sets, items, and relations. Moreover, the operation is associated to a trigger widget that fires its execution. Finally, the operation causes a state transition on the interface, which is associated to a change of contextual information, such as the exploration trail, and layout reconfigurations. These are some abstract interface elements identified along the research.

Another future work is investigating how exploration patterns emerge within a community of information consumers and how to efficiently support the generalization and reuse of exploration strategies. One obstacle for such user study is its long-term nature, which depends on the continuous use of the environment among a specific community of users.

Finally, the formal description of exploration actions also opens a range of possibilities for research on intelligent computational assistance to exploration tasks, where machine-learning agents can be trained to recommend exploration actions and strategies based on previously recorded explorations.

## Acknowledgements

The authors were partially supported by CNPq project 557128/200-9 National Science, Technology Institute on Web Science, CAPES, Instituto Federal Fluminense, and Google Research Program.

## REFERENCES

[1] Cc Aggarwal and Haixun Wang. 2010. *Managing and mining graph data*. Springer US.

[2] Fahad Alahmari, James A Thom, Liam Magee, and Wilson Wong. 2012. Evaluating Semantic Browsers for Consuming Linked Data. In *Proceedings of the Twenty-Third Australasian Database Conference - Volume 124*, 89–98. Retrieved from http://dl.acm.org/citation.cfm?id=2483751

[3] Robert a. Amar and John T. Stasko. 2005. Knowledge precepts for design and evaluation of information visualizations. *IEEE Trans. Vis. Comput. Graph.* 11, 4 (2005), 432–442. DOI:https://doi.org/10.1109/TVCG.2005.63

[4] Robert Amar, James Eagan, and John Stasko. 2005. Low-level components of analytic activity in information visualization. In *IEEE Symposium on Information Visualization, 2005. INFOVIS 2005.*, 111–117. DOI:https://doi.org/10.1109/INFVIS.2005.1532136

[5] Samur Araujo, Geert-Jan Houben, Daniel Schwabe, and Jan Hidders. 2010. Fusion - Visually Exploring and Eliciting Relationships in Linked Data. In *Proceedings of the 9th International Semantic Web Conference on The Semantic Web - Volume Part I* (ISWC'10), 1–15. DOI:https://doi.org/10.1007/978-3-642-17746-0_1

[6] Samur Araújo and Daniel Schwabe. 2009. Explorator: a tool for exploring RDF data through direct manipulation. *Linked data web* (2009). Retrieved from http://ceur-ws.org/Vol-538/ldow2009_paper2.pdf

[7] Marcelo Arenas. 2014. SemFacet : Semantic Faceted Search over Yago *. (2014), 3–6. DOI:https://doi.org/10.1109/2567948.2577011

[8] Marcia J. Bates. 1989. The design of browsing and berrypicking techniques for the online search interface. *Online Inf. Rev.* 13, 5 (1989), 407–424. DOI:https://doi.org/10.1108/eb024320

[9] Richard Bird and Philip Wadler. 1988. *An Introduction to Functional Programming*. Prentice Hall International (UK) Ltd., Hertfordshire, UK, UK.

[10] A Bozzon, M Brambilla, S Ceri, and D Mazza. 2013. Exploratory search framework for Web data sources. *VLDB J.* 22, 5 (2013), 641–663.

[11] Alessandro Bozzon, Marco Brambilla, Stefano Ceri, and Piero Fraternali. 2010. Liquid Query: Multi-Domain Exploratory Search on the Web. In *Proceedings of the 19th international conference on World wide web - WWW '10*, 161. DOI:https://doi.org/10.1145/1772690.1772708

[12] Matthew Brehmer and Tamara Munzner. 2013. A multi-level typology of abstract visualization tasks. *IEEE Trans. Vis. Comput. Graph.* 19, 12 (2013), 2376–2385. DOI:https://doi.org/10.1109/TVCG.2013.124

[13] Sören Brunk and Philipp Heim. 2011. Tfacet: Hierarchical faceted exploration of semantic data using well-known interaction concepts. *CEUR Workshop Proc.* 817, (2011), 31–36.

[14] Sven Buschbeck, Anthony Jameson, Adrian Spirescu, Tanja Schneeberger, Raphaël Troncy, Houda Khrouf, Osma Suominen, and Eero Hyvönen. 2013. Parallel faceted browsing. In *CHI '13 Extended*

*Abstracts on Human Factors in Computing Systems on - CHI EA '13*, 3023. DOI:https://doi.org/10.1145/2468356.2479601

[15]     Rick Cattell. 2011. Scalable SQL and NoSQL data stores. *ACM SIGMOD Rec.* 39, 4 (2011), 12. DOI:https://doi.org/10.1145/1978915.1978919

[16]     PPS Chen. 1976. The entity-relationship model—toward a unified view of data. *ACM Trans. Database Syst.* 1, 1 (1976), 9–36. DOI:https://doi.org/10.1145/320434.320440

[17]     Ed H Chi. 2000. A Taxonomy of Visualization Techniques using the Data State Reference Model. *Inf. Vis. 2000. InfoVis 2000. IEEE Symp.* 94301, Table 2 (2000), 69–75. DOI:https://doi.org/10.1109/INFVIS.2000.885092

[18]     Marcelo Cohen and Daniel Schwabe. 2012. Support for Reusable Explorations of Linked Data in the Semantic Web. In *Current Trends in Web Engineering*, Nora Harth, Andreas and Koch (ed.). Springer Berlin Heidelberg, 119–126. DOI:https://doi.org/10.1007/978-3-642-27997-3_11

[19]     Vania Dimitrova, Lydia Lau, Dhavalkumar Thakker, Fan Yang-Turner, and Dimoklis Despotakis. 2013. Exploring exploratory search: a user study with linked semantic data. In *Proceedings of the 2nd International Workshop on Intelligent Exploration of Semantic Data - IESD '13*, 1–8. DOI:https://doi.org/10.1145/2462197.2462199

[20]     S. Ferré and a. Hermann. 2012. Reconciling faceted search and query languages for the Semantic Web. *Int. J. Metadata, Semant. Ontol.* 7, 1 (2012), 37.

[21]     Roberto García, Josep Maria Brunetti, Rosa Gil, and Juan Manuel Gimeno. 2013. Rhizomer : Overview, Facets and Pivoting for Semantic Data Exploration. In *International Workshop on Intelligent Exploration of Semantic Data*.

[22]     Jeffrey Heer, Jock D. Mackinlay, Chris Stolte, and Maneesh Agrawala. 2008. Graphical histories for visualization: Supporting analysis, communication, and evaluation. *IEEE Trans. Vis. Comput. Graph.* 14, 6 (2008), 1189–1196. DOI:https://doi.org/10.1109/TVCG.2008.137

[23]     Philipp Heim, Steffen Lohmann, and Timo Stegemann. 2010. Interactive Relationship Discovery via the Semantic Web. *Proc. 7th Ext. Semant. Web Conf. (ESWC 2010)* 6088, C (2010), 303–317.

[24]     Philipp Heim, Jürgen Ziegler, and Steffen Lohmann. 2008. gFacet : A Browser for the Web of Data. In *Proceedings of the International Workshop on Interacting with Multimedia Content in the Social Semantic Web*, 49–58. DOI:https://doi.org/10.1.1.143.1026

[25]     Michiel Hildebrand, Jacco van Ossenbruggen, and Lynda Hardman. 2006. /facet: A Browser for Heterogeneous Semantic Web Repositories. In *International Semantic Web Conference*, 272–285. DOI:https://doi.org/10.1007/11926078_20

[26]     Angelo Di Iorio, Raffaele Giannella, Francesco Poggi, and Fabio Vitali. 2015. Exploring Bibliographies for Research-related Tasks. In *Proceedings of the 24th International Conference on World Wide Web - WWW '15 Companion*, 1001–1006. DOI:https://doi.org/10.1145/2740908.2742018

[27]     Carol C Kuhlthau. 1991. Inside the search process: Information seeking from the user's perspective. *J. Am. Soc. Inf. Sci.* 42, 5 (June 1991), 361–371. DOI:https://doi.org/10.1002/(SICI)1097-4571(199106)42:5<361::AID-ASI6>3.0.CO;2-#

[28]     Gary Marchionini. 2006. From finding to understanding. *Commun. ACM* 49, 4 (2006), 41–46.

[29]     Timo Niemi and Kalervo Järvelin. 1983. A straightforward formalization of the relational model. *ACM SIGMOD Record 14*, 15–38. DOI:https://doi.org/10.1145/984540.984542

[30]     Eyal Oren, Renaud Delbru, and Stefan Decker. 2006. Extending faceted navigation for RDF data. *Int. Semant. web Conf.* 4273, (2006), 559–572. DOI:https://doi.org/10.1007/11926078

[31]     Silvio Peroni, Alexander Dutton, Tanya Gray, and David Shotton. 2015. Setting our bibliographic references free: towards open citation data. *J. Doc.* 71, 2 (March 2015), 253–277. DOI:https://doi.org/10.1108/JD-12-2013-0166

[32]    Igor O. Popov, M. C. Schraefel, Wendy Hall, and Nigel Shadbolt. 2011. Connecting the Dots: A Multi-pivot Approach to Data Exploration. In *Proceedings of The International Semantic Web Conference (ISWC 2011)*, 553–568. DOI:https://doi.org/10.1007/978-3-642-25073-6_35

[33]    Martin Przyjaciel-Zablocki, Alexander Schätzle, Thomas Hornung, and Georg Lausen. 2011. Rdfpath: Path query processing on large rdf graphs with mapreduce. *Eswc* (2011), 50–64.

[34]    m. c. schraefel, Daniel A Smith, Alisdair Owens, Alistair Russell, Craig Harris, and Max Wilson. 2005. The evolving mSpace platform. In *Proceedings of the sixteenth ACM conference on Hypertext and hypermedia - HYPERTEXT '05*, 174. DOI:https://doi.org/10.1145/1083356.1083391

[35]    B. Shneiderman. 1996. The eyes have it: a task by data type taxonomy for information visualizations. In *Proceedings 1996 IEEE Symposium on Visual Languages*, 336–343. DOI:https://doi.org/10.1109/VL.1996.545307

[36]    Amit Singhal. 2012. Introducing the Knowledge Graph: things, not strings. *Inside Search Blog*. Retrieved September 17, 2017 from http://googleblog.blogspot.com.br/2012/05/introducing-knowledge-graph-things-not.html

[37]    Yannis Tzitzikas, Nikos Manolis, and Panagiotis Papadakos. 2017. Faceted exploration of RDF/S datasets: a survey. *J. Intell. Inf. Syst.* 48, 2 (April 2017), 329–364. DOI:https://doi.org/10.1007/s10844-016-0413-8

[38]    Ryen W. White and Resa A. Roth. 2009. Exploratory Search: Beyond the Query-Response Paradigm. *Synth. Lect. Inf. Concepts, Retrieval, Serv.* 1, 1 (January 2009), 1–98. DOI:https://doi.org/10.2200/S00174ED1V01Y200901ICR003

[39]    Ryen W White, Steven M. Drucker, Gary Marchionini, Marti Hearst, and m. c. Schraefel. 2007. Exploratory Search and HCI: Designing and Evaluating Interfaces to Support Exploratory Search Interaction. In *CHI '07 extended abstracts on Human factors in computing systems - CHI '07*, 2877. DOI:https://doi.org/10.1145/1240866.1241100

[40]    Barbara M. Wildemuth and Luanne Freund. 2012. Assigning search tasks designed to elicit exploratory search behaviors. *Proc. Symp. Human-Computer Interact. Inf. Retr. - HCIR '12* C (2012), 1–10. Retrieved from http://dl.acm.org/citation.cfm?id=2391224.2391228

[41]    Max L. Wilson, M.c. Schraefel, and Ryen W. White. 2009. Evaluating advanced search interfaces using established information-seeking models. *J. Am. Soc. Inf. Sci. Technol.* 60, 7 (July 2009), 1407–1422. DOI:https://doi.org/10.1002/asi.21080

[42]    Sivan Yogev, Haggai Roitman, David Carmel, and Naama Zwerdling. 2012. Towards expressive exploratory search over entity-relationship data. In *Proceedings of the 21st international conference companion on World Wide Web - WWW '12 Companion*, 83. DOI:https://doi.org/10.1145/2187980.2187990

[43]    Junliang Zhang and Gary Marchionini. 2005. Evaluation and Evolution of a Browse and Search Interface: Relation Browser++. *6th Annu. Natl. Conf. Digit. Gov. Res. (dg.o 2005)* January (2005), 179–188. DOI:https://doi.org/10.1145/1065226.1065279